\documentclass[preprint]{article}
\newcommand{\be}{\begin{equation}} \newcommand{\ee}{\end{equation}} 
\newcommand{\bea}{\begin{eqnarray}}\newcommand{\eea}{\end{eqnarray}}

\begin{document}
\title{ Deconstructing non-Dirac-hermitian supersymmetric quantum systems}
\author{ Pijush K. Ghosh}
\date{Department of Physics, Siksha-Bhavana,\\ 
Visva-Bharati University,\\
Santiniketan, PIN 731 235, India.}
\maketitle
\begin{abstract} 
A method to construct non-Dirac-hermitian supersymmetric quantum
system that is isospectral with a Dirac-hermitian Hamiltonian is
presented. The general technique involves a realization of the basic
canonical (anti-)commutation relations involving both bosonic and fermionic
degrees of freedom in terms of non-Dirac-hermitian operators which are
hermitian in a Hilbert space that is endowed with a pre-determined
positive-definite metric. A pseudo-hermitian realization of the Clifford
algebra for a pre-determined positive-definite metric is used to construct
supersymmetric systems with one or many degrees of freedom. It is shown that
exactly solvable non-Dirac-hermitian supersymmetric quantum systems can be
constructed corresponding to each exactly solvable Dirac-hermitian system.
Examples of non-Dirac-hermitian (i) non-relativistic Pauli Hamiltonian,
(ii) super-conformal quantum system and (iii) supersymmetric Calogero-type
models admitting entirely real spectra are presented.
\end{abstract}
\tableofcontents{}

\section{Introduction}

The study of non-Dirac-hermitian quantum systems with unbroken combined
parity(${\cal{P}}$) and time-reversal(${\cal{T}}$) symmetry has received
considerable attention over the last few years\cite{bend,ali,quasi,mostafa1,
quesne,quesne3,quesne2, pani, andrianov,piju1,ijtp,me,piju,ptcsm,pkumar,europe}.
A consistent quantum description
of such systems, including reality of the entire spectra and unitary
time-evolution, is admissible with the choice of a new inner-product
in the Hilbert space\cite{bend}. A non-Dirac-hermitian quantum system
admitting entirely real spectra may also be understood in terms of
pseudo-hermitian operators\cite{ali,quasi}. The existence of a
positive-definite metric in the Hilbert space is again crucial in this
formalism for showing reality of the entire spectra as well as unitary
time evolution.  Supersymmetric versions of ${\cal{PT}}$ symmetric and/or
pseudo-hermitian quantum systems have been studied in the
literature\cite{mostafa1,quesne,quesne3, quesne2, pani, andrianov,me}.

A general problem encountered in the study of ${\cal{PT}}$ symmetric
non-Dirac-hermitian quantum system is the construction of the metric in
the Hilbert space, which is essential for the calculation of expectation
values of physical observables as well as different correlation functions.
The description of a quantum system without the knowledge of the metric
is thus incomplete, even though the complete spectrum and the associated
eigenfunctions may be be known explicitly. An approach taken in
Ref. \cite{piju1} was to consider Hilbert space with a pre-determined metric
so that non-Dirac-hermitian quantum system can be constructed from known
Dirac-hermitian Hamiltonian through isospectral deformation. Although the
non-Dirac-hermitian Hamiltonian constructed in this way is isospectral with
the corresponding Dirac-hermitian Hamiltonian, the difference may appear in
the description of different correlation functions of these two quantum
systems\cite{piju}.

Several exactly solvable non-Dirac-hermitian quantum systems with a complete
description were constructed following this approach\cite{piju1}. It is
reassuring that the asymmetric $XXZ$ spin-chain\cite{xxz}, which has been
studied extensively in the context of two species reaction-diffusion processes
and Kardar-Parisi-Zhang-type growth phenomenon, was shown to be
pseudo-hermitian following this general approach\cite{piju1}. A
non-Dirac-hermitian transverse-field Ising model appears as a special case
of this general class of spin systems\cite{piju}. Further, the celebrated Dicke
model\cite{dicke} was shown to admit bound states for previously unexplored
range of parameters\cite{me}. These models may be considered as prototype
examples of non-Dirac-hermitian quantum systems with a complete description
which provide the ground for testing any idea related to the subject,
including validity of any approximate or numerical method.

The purpose of this article is to generalize the approach of Ref. \cite{piju1}
to include fermionic degrees of freedom and construct non-Dirac-hermitian
supersymmetric quantum systems with a pre-determined metric in the Hilbert
space. The general method involves a realization of the basic canonical
(anti-)commutation relations involving both bosonic and fermionic degrees
of freedom
in terms of non-Dirac-hermitian operators which are hermitian with respect to
a modified inner-product the Hilbert space. It may be noted that the Hilbert
space of a supersymmetric system is $Z_2$-graded and the metric can be
expressed as a  direct-product of metrics corresponding to bosonic and fermionic
degrees of freedom. The method described in Ref. \cite{piju1} is valid for the
purely bosonic sector of a supersymmetric system which involves canonical
relations involving only bosonic degrees of freedom. A pseudo-hermitian
realization of the Clifford algebra for a pre-determined positive-definite
metric is given in this article. A general construction of pseudo-hermitian
supersymmetric systems with one or many degrees of freedom is presented.
Further, it is shown that exactly solvable pseudo-hermitian supersymmetric
quantum systems can be constructed corresponding to each exactly solvable
Dirac-hermitian system. Examples of pseudo-hermitian (i) Pauli
Hamiltonian, (ii) super-conformal quantum system\cite{dff,fr,cmm} and
(iii) Calogero-type models\cite{csm,poly,pcalo,sae}
are also presented.

The article is organized as follows. A pseudo-hermitian realization of
the Pauli matrices is presented in the next section. A generalization
of these results to the Clifford algebra with an arbitrary number of
elements is given in section 3. The section 4 contains discussions on
pseudo-hermitian supersymmetric quantum systems with one degree of freedom.
It is shown that exactly solvable pseudo-hermitian quantum system can be
constructed corresponding to each Dirac-hermitian system with a shape-invariant
potential. A general formulation of pseudo-hermitian supersymmetric quantum
system with many degrees of freedom is presented in section 5. Examples of 
pseudo-hermitian Pauli Hamiltonian, super-conformal quantum system and
Calogero-type models are given in sections 6.1, 6.2 and 6.3, respectively.
Finally, the results are summarized in section 7.

\section{Pauli Matrices: A non-Dirac-hermitian realization}

The Pauli matrices are hermitian with the standard inner-product
$\langle \cdot | \cdot \rangle$ in the space of its eigen-vectors
${\cal{H}}_D$. A vector-space ${{\cal{H}}_{\eta_+^f}}$ that is endowed
with the positive-definite metric $\eta_+^f$,
\be
\eta_+^f := exp \left ( - \vec{\sigma} \cdot \hat{n} \phi \right ),
\ \ \hat{n} \cdot \hat{n} = 1, \ (\phi, n^a) \in R, \ \ a=1, 2, 3,
\ee
\noindent and the inner-product
$\langle \langle \cdot | \cdot \rangle \rangle_{{\cal{H}}_{\eta_+^f}}
= \langle \cdot | \eta_+^f \cdot \rangle$ is introduced. The positivity
of the metric follows from the fact that the eigen values of the matrix
$\vec{\sigma} \cdot \hat{n}$ are real. 
With the introduction of the similarity operator $\rho^f$, 
\be
\rho^f := \sqrt{\eta_+^f}=exp \left ( - \frac{\vec{\sigma} \cdot \hat{n}}{2}
\phi \right ),
\ee
\noindent a set of non-Dirac-hermitian matrices may be introduced in terms of
the Pauli matrices $\sigma^a$ as follows:
\bea
\Sigma^a & := & (\rho^f)^{-1} \sigma^a \rho^f\nonumber \\
& = & \sum_{b=1}^3 R^{ab} \sigma^b\nonumber \\
R^{ab} & \equiv & n^a n^b \left ( 1 - cosh \phi \right ) +
\delta^{ab} cosh \phi + i \epsilon^{abc} n^c sinh \phi.
\eea
\noindent Note that $R_{ij}^* \neq R_{ij}$ and $\sum_{j=1}^3 R_{ij}^2 =1
\forall i$. The matrices $\Sigma^a$ obey the same algebra satisfied by the
Pauli matrices:
\bea
&& \left [ \Sigma^a, \Sigma^b \right ] = 2 i \epsilon^{abc} \Sigma^c, \ \
\left \{ \Sigma^a, \Sigma^b \right \} = 2 \delta^{ab}\nonumber \\
&& \Sigma_{\pm} := \frac{1}{2} \left ( \Sigma^1 \pm i \Sigma^2 \right ), \ \
\{\Sigma_-, \Sigma_+ \} = 1, \ \ \Sigma_{\pm}^2 = 0.
\eea
\noindent and are hermitian in the vector space ${{\cal{H}}_{\eta_+^f}}$.

A few comments are in order at this point.\\
(i) The matrices $\Sigma^a$ depend on three real independent parameters.
A general $2 \times 2$ non-Dirac-hermitian matrix may be constructed in terms
of $\Sigma^a$'s and the $2 \times 2$ identity matrix $I$,
\be
\Sigma = p^0 I + \sum_{a=1}^3 p^a \Sigma^a, \ \ (p^0, p^a) \in R,
\ee
\noindent where $p^0$ and $p^a$ are four real parameters. The matrix
depends on seven independent real parameters and is hermitian in the
vector space ${{\cal{H}}_{\eta_+^f}}$. The real eigenvalues $\lambda_{\pm}$
and the associated eigenvectors $v_{\pm}$ of $\Sigma^a$'s are,
\bea
&&\lambda_{\pm}=p^0 \pm \sqrt{p}, \ \ p \equiv
\sqrt{\sum_{a=1}^3 (p^a)^2},\nonumber \\
&& v_{\pm} = N_{\pm} \left ( \rho^f \right )^{-1}
\pmatrix{ {\lambda_{\pm} + p^3 -p^0}\cr {p^1+ip^2}}, \ \
N_{\pm} \equiv \left [ 2 \left ( p^2 \pm p^3 p \right ) \right ]^{-
\frac{1}{2}}.
\eea
\noindent The most general $2 \times 2$ matrix with complex elements depends
on eight real parameters. The complex matrix $\Sigma$ admitting entirely real
eigenvalues depends on seven real parameters and thus, is more general
than the one presented in Ref. \cite{sing}. Moreover, the method in
constructing $\Sigma$ is completely different from the one followed 
previously\cite{sing}. The matrix $\Sigma$ may be used as the toy-model for
studying different ideas related to ${\cal{PT}}$ symmetric and pseudo-hermitian
quantum systems.

(ii) Pseudo-hermitian spin chain systems and Dicke models have been constructed
previously\cite{me,piju1,piju} by using $\Sigma^a$'s with the choice
$n^1=n^2=0,n^3=1$. More general pseudo-hermitian spin chain systems and Dicke
models may be constructed by using $\Sigma^a$'s with arbitrary $n^1,n^2,n^3$.

(iii) Any Dirac-hermitian representation of the Pauli matrices is unitary
equivalent to the standard representation, where $\sigma^3$ is taken to be
diagonal. The unitary transformations that relate different representations 
correspond to gauge transformations and within the formalism of hermitian
matrix models, real gauge potentials are constructed using the unitary matrix.
On the other hand, the non-Dirac-hermitian matrices $\Sigma^a$'s are equivalent
to the Pauli matrices through a non-unitary similarity transformation. Such
transformations can again be interpreted as gauge transformations with complex
gauge potentials.

(iv) The pseudo-hermitian matrix $\Sigma$ may be used to construct
seven-parameter dependent $2 \times 2$ pseudo-unitary matrix $D$ and
its inverse $D^{-1}$,
\be
D = e^{i \Sigma}, \ \ D^{-1} = e^{-i \Sigma}.
\ee
\noindent The inner-product
$\langle \langle v, u \rangle \rangle_{{\cal{H}}_{\eta_{+}}^f}$ involving
two arbitrary vectors $u, v$ remains invariant under the pseudo-unitary
transformation $ u \rightarrow u^{\prime}=D u, v \rightarrow v^{\prime}=D v $,
which can be shown using the relation\cite{jain},
\be
D^{\dagger} = \eta_+^f D (\eta_+^f)^{-1}.
\ee
\noindent Such pseudo-unitary matrices may have applications in the study
of pseudo-hermitian random matrix model\cite{jain}.

(v) An anti-linear ${\cal{PT}}$ transformation may be introduced
with the actions of the anti-linear operator ${\cal{T}}$ and the discrete
symmetry operator ${\cal{P}}$ on $\sigma^1, \sigma^2,\sigma^3$ as follows:
\bea
{\cal{T}}: && i \rightarrow - i, \ \sigma^a \rightarrow \sigma^a \
\forall a;\nonumber \\
{\cal{P}}: && \sigma^1 \rightarrow \tilde{\sigma}^1 = \sigma^1 cos \beta +
\sigma^2 sin \beta,\nonumber \\
&&\sigma^2 \rightarrow \tilde{\sigma}^2 = \sigma^1 sin \beta -
\sigma^2 cos \beta,\nonumber \\
&& \sigma^3 \rightarrow \tilde{\sigma}^3 = \sigma^3, \ \
0 \leq \beta \leq 2 \pi.
\label{pteqs}
\eea
\noindent It may be checked that $\Sigma^3$, which appears in the
description of single-particle supersymmetric non-Dirac-hermitian
Hamiltonian in section 4, is ${\cal{PT}}$ symmetric for $\theta= 2 tan^{-1}
\frac{\eta^2}{\eta^1}$.

\section{A pseudo-hermitian realization of Clifford algebra}

The real Clifford algebra with $2N$ elements satisfy the relations,
\be
\{\xi_p, \xi_q \} = 2 \delta_{pq}, \ \ p, q=1, 2, \dots, 2N.
\ee
\noindent The complexification of the algebra can be achieved by
introducing the fermionic variables $\psi_i$ and their conjugates
$\psi_i^{\dagger}$ in ${\cal{H}}_D$,
\be
\psi_i:= \frac{1}{2} \left ( \xi_i - i \xi_{N+i} \right ), \ \
\psi_i^{\dagger}:= \frac{1}{2} \left (\xi_i + i \xi_{N+i} \right ), \ \
i, j=1, 2, \dots N.
\label{fer}
\ee
\noindent These fermionic variables satisfy the complex Clifford algebra,
\be
\{\psi_i,\psi_j\}=0=\{\psi_i^{\dagger},\psi_j^{\dagger}\}, \ \
\{\psi_i, \psi_j^{\dagger}\}=\delta_{ij}, 
\ee
\noindent and either $\xi_p$ or $\psi_i, \psi_i^{\dagger}$ may be used
to construct Dirac-hermitian supersymmetric systems.

A realization of the $N (2 N-1)$ number of generators of the $O(2N)$ group
is given in terms of elements of the Clifford algebra $\xi_p$ as follows,
\be
J_{pq} := \frac{i}{4} \left [ \xi_p, \xi_q \right ].
\ee
\noindent These generators may be used to obtain a multi-parameter dependent
pseudo-hermitian realization of the Clifford algebra, much akin to the case
of Pauli matrices. A `complex rotation' in the space of elements $\xi_p$ is
described in terms of the hermitian operator $\eta_+^f$,
\be
\eta_+^f := e^{-T}, \ \ T:= \frac{1}{2} \sum_{p, q=1}^{2N} t_{pq} J_{pq} , \ \ 
t_{pq}=-t_{qp}, 
\ee
\noindent where $t_{pq}$ are $N (2 N-1)$ real parameters. The
positive-definiteness of $\eta_+^f$ follows from the fact that the eigen
values of $T$ are real in the whole of the parameter space. It may be
noted that the operator $T$ can be expressed as a quadratic form of fermionic
operators $\psi_i(\psi_i^{\dagger})$ which is known\cite{lsm,hemen} to admit
entirely real spectra. Without loss of any generality, a particular form of
$\eta_+^f$ is chosen in this article for its simplicity,
\be
\eta_+^f := \prod_{i=1}^N e^{2 \gamma_i J_{i N+i}-\gamma_i}, \ \
\rho^f := \sqrt{\eta_+^f}=\prod_{i=1}^N e^{ \gamma_i J_{i N+i}-
\frac{\gamma_i}{2}},
\ \ \gamma_i \in R \ \forall \ i.
\label{ord}
\ee
\noindent It may be noted that ordering of the generators $J_{iN+i}$ is not
required in Eq.(\ref{ord}), since the commutators $[J_{iN+i}, J_{jN+j}]=0$
for any $i$ and $j$. A set of non-Dirac-hermitian elements of the
real Clifford algebra is introduced as follows, 
\bea
\Gamma_p & := & (\rho^f)^{-1} \xi_p \rho^f,\nonumber \\
\Gamma_ i & = & \xi_i cosh \gamma_i 
+ i \xi_{N+i} sinh \gamma_i,\nonumber \\
\Gamma_{N+i} & = & - i \xi_i sinh \gamma_i + \xi_{N+i} cosh \gamma_i.
\eea
\noindent The elements $\Gamma_p$ are hermitian in ${\cal{H}}_{\eta_+}$.
A pseudo-hermitian realization of the generators of the group $O(2N+1)$
is facilitated by the introduction of the element $\Gamma_{2N+1}$,
\be
\Gamma_{2N+1} := \Gamma_1 \Gamma_2 \dots \Gamma_{2N-1} \Gamma_{2N}=
\xi_1 \xi_2 \dots \xi_{2N-1} \xi_{2N},
\ee
\noindent which anti-commutes with all the $\Gamma_p/\xi_p$'s and squares
to unity.

A set of fermionic operators $\Psi_i$'s and their adjoints $\Psi_i^{\dagger}$
in ${\cal{H}}_{\eta_+}$ may be defined in terms of $\Gamma_p$
as,
\bea
\Psi_i & := & \frac{1}{2} \left ( \Gamma_i - i \Gamma_{N+i} \right )
= e^{-\gamma_i} \psi_i,\nonumber \\
\Psi_i^{\dagger} & := & \frac{1}{2} \left ( \Gamma_i +
i \Gamma_{N+i} \right)= e^{\gamma_i} \psi_i^{\dagger},
\eea
\noindent which satisfy the basic canonical anti-commutation relations.
\be
\{\Psi_i,\Psi_j\}=0=\{\Psi_i^{\dagger},\Psi_j^{\dagger}\}, \ \
\{\Psi_i, \Psi_j^{\dagger}\}=\delta_{ij}.
\ee
\noindent The metric $\eta_+^f$ and the similarity operator $\rho^f$ are
expressed in terms of $\psi_i, \psi_i^{\dagger}$ as,
\be
\eta_+^f = \prod_{i=1}^N e^{- 2 \gamma_i \psi_i^{\dagger} \psi_i} 
 = \prod_{i=1}^N e^{- 2 \gamma_i \Psi_i^{\dagger} \Psi_i} 
\label{met}
\ee
\noindent The total fermion number operator $N_f$ has identical expression,
\be
N_f=\sum_{i=1}^N \psi_i^{\dagger} \psi_i=
\sum_{i=1}^N \Psi_i^{\dagger} \Psi_i,
\ee
\noindent in ${\cal{H}}_D$ as well as in
${\cal{H}}_{\eta_+}$. However, a general eigenstate 
$|f_1, \dots, f_i, \dots, f_N\rangle_{{\cal{H}}_D}$ of $N_f$ in ${\cal{H}}_D$,
is related to the
corresponding state $|f_1, \dots, f_i, \dots, f_N\rangle_{{\cal{H}}_{\eta_+}}$
in the Hilbert space ${\cal{H}}_{\eta_+}$ through the following relation:
\be
|f_1, \dots, f_i, \dots, f_N\rangle_{{\cal{H}}_{\eta_+}} = \prod_{k=1}^N
e^{\gamma_k f_k} |f_1, \dots, f_i, \dots, f_N\rangle_{{\cal{H}}_D}, \ \
f_i = 0, 1 \ \forall \ i.
\ee
\noindent The $2^N$ states
$|f_1, \dots, f_i, \dots, f_N\rangle_{{\cal{H}}_{\eta_+}}$
form a complete set of orthonormal states in ${\cal{H}}_{\eta_+}$,
while $|f_1, \dots, f_i, \dots, f_N\rangle_{{\cal{H}}_{D}}$ constitute a
complete set of orthonormal states in ${\cal{H}}_D$. The action of
$\Psi_i(\Psi_i^{\dagger})$ on
$|f_1, \dots, f_i, \dots, f_N\rangle_{{\cal{H}}_{\eta_+}}$
is identical to that of $\psi_i(\psi_i^{\dagger})$ on
$|f_1, \dots, f_i, \dots, f_N\rangle_{{\cal{H}}_D}$. In particular,
\bea
\Psi_i |f_1, \dots, f_i, \dots f_N\rangle_{{\cal{H}}_{\eta_+}} & = & 0,
\ \ \ \
if \ \ \ \ f_i=0,\nonumber \\
& = & |f_1, \dots, 0, \dots f_N\rangle_{{\cal{H}}_{\eta_+}}, \ \ \ \
if \ \ \ \ f_i=1,\nonumber \\
\Psi_i^{\dagger} |f_1, \dots, f_i, \dots f_N\rangle_{{\cal{H}}_{\eta_+}}
& = & 0, \ \ \ \ if \ \ \ \ f_i=1,\nonumber \\
& = & |f_1, \dots, 1, \dots f_N\rangle_{{\cal{H}}_{\eta_+}}, \ \ \ \
if \ \ \ \ f_i=0.
\eea
\noindent Either $\Gamma_i$ or $\Psi_i, \Psi_i^{\dagger}$ may be used to
construct pseudo-hermitian supersymmetric quantum systems.

The known representation\cite{clifford,ritten} of the elements $\xi_p$ in
terms of $2^N \times 2^N$ matrices can be used to find the corresponding
representation for $\Gamma_p, \Psi_i, \Psi_i^{\dagger}$. In general, the
matrices $\Gamma_p$ depend on $N$ real parameters $\gamma_i$. A $2^{N} \times
2^{N}$ pseudo-hermitian matrix depending on $2^{2N} + N$ real parameters
may be constructed in terms of $\Gamma_p$'s and the $2^{N} \times 2^{N}$
identity matrix $I$ as,
\bea
\Gamma & = & a^0 I + \sum_{p=1}^{2N} a_p^1 \Gamma_p + \sum_{p_1< p_2=1}^{2N}
a_{p_1p_2}^2 \Gamma_{p_1} \Gamma_{p_2} + \dots\nonumber \\
& + & \sum_{p_1<p_2 \dots <p_j=1}^{2N}
a_{p_1p_2 \dots p_j}^j \Gamma_{p_1} \Gamma_{p_2} \dots \Gamma_{p_j} + \dots
+ a^{2N} \Gamma_1 \Gamma_2 \dots \Gamma_{2N}
\label{gamma}
\eea
\noindent where $a^0, a_{p}^1, a_{pq}^2 \dots, a^{2N}$ are $2^{2N}$ real
parameters.
The eigenvalues of the matrix $\Gamma$ are real by construction. A complete
set of orthonormal eigen-vectors can also be constructed in
${\cal{H}}_{\eta_+}$.

It appears that $\Gamma$ is not the most general
$2^N \times 2^N$ pseudo-hermitian matrix having entirely real eigen-values.
For example, a non-Dirac-hermitian matrix $\tilde{\Gamma}$ depending on
$2^{2N} + 3 N$ real parameters may be constructed by replacing
$\Gamma_p \rightarrow \tilde{\Gamma}_p$ in Eq. (\ref{gamma}), where
\bea
\tilde{\Gamma}_p & := & (\tilde{\rho}^f)^{-1} \xi_p \tilde{\rho}^f\nonumber \\
\tilde{\rho}^f & := & \prod_{i=1}^N \otimes \ exp \left ( -
\frac{\vec{\sigma} \cdot \hat{n}_i}{2} \phi_i \right ), \ \
\hat{n}_i \cdot \hat{n}_i = 1 \ \forall i,
\eea
\noindent and it is understood that $\xi_p$'s are realized in terms of Pauli
matrices\cite{clifford,ritten}. The matrix $\tilde{\Gamma}$ thus constructed will have
entirely real eigenvalues with a complete set of orthonormal vectors in
${\cal{H}}_{\eta_+}$. Other possibilities including a more general operator
$T$ appearing in the definition of $\eta_+^f$ and with more number of real
parameters also exist, which will not be pursued in this article.

A comment regarding the ${\cal{PT}}$ symmetry in the space of the elements
of the Clifford algebra is in order before the end of this section.
An anti-linear ${\cal{PT}}$ transformation may be introduced with the action
of the anti-linear operator ${\cal{T}}$ and the discrete symmetry${\cal{P}}$
on the elements $\xi_p$ as,
\bea
{\cal{T}}: && i \rightarrow -i, \ \xi_p \rightarrow \xi_p \ \forall
p;\nonumber \\
{\cal{P}}: && \xi_i \rightarrow \tilde{\xi}_i = \xi_i cos \beta +
\xi_{N+i} sin \beta,\nonumber \\ 
&& \xi_{N+i} \rightarrow \tilde{\xi}_{N+i} = \xi_i sin \beta -
\xi_{N+i} cos \beta, \ \ 0 \leq \beta \leq 2 \pi.
\eea
\noindent The action of the ${\cal{PT}}$ transformation on the
fermionic variables $\psi_i$ is as follows,
\bea
&& {\cal{P}}: \psi_i \rightarrow e^{-i \beta} \psi_i^{\dagger}, \ \
\psi_i^{\dagger} \rightarrow e^{ i \beta} \psi_i,\nonumber \\
&& {\cal{T}}: \psi_i \rightarrow \psi_i^{\dagger}, \ \
\psi_i^{\dagger} \rightarrow \psi_i\nonumber \\
&& {\cal{PT}}: \psi_i \rightarrow e^{i \beta} \psi_i, \ \
\psi_i^{\dagger} \rightarrow e^{-i \beta} \psi_i^{\dagger}, \
\label{pteqf}
\eea
\noindent where $\beta$ appears as a phase which may be fixed at
some specific value depending on the physical requirements. A
supersymmetric Hamiltonian in the linear realization of the super-algebra
contains bilinear terms involving fermionic variables. It may be noted
that the bilinear terms of the form $\psi_i^{\dagger} \psi_j$ are
${\cal{PT}}$ invariant for any $\beta$. However, bilinear terms of the
form $\psi_i \psi_j, \psi_i^{\dagger} \psi_j^{\dagger}$ are ${\cal{PT}}$
invariant for $\beta=0$. A particular nonlinear realization\cite{pcalo}
of the super-algebra involves $\Gamma_{2N+1}$ which is also ${\cal{PT}}$
invariant for any $\theta$. This can be checked by expressing
$\Gamma_{2N+1}$  in terms of fermionic variables as,
\be
\Gamma_{2N+1} = (-1)^N \prod_{i=1}^N \left ( 2 \psi_i^{\dagger} \psi_i
-1 \right ).
\ee
\noindent This provides a framework for constructing ${\cal{PT}}$ symmetric
supersymmetric non-Dirac-hermitian Hamiltonian.

\section{Single-particle non-Dirac-hermitian SUSY}

The supercharges are introduced as follows:
\bea
Q_1 & = & \left [ p + i W_1^{\prime}(x) \right ] \Sigma_-,\nonumber \\
Q_2 & = & \left [ p - i W_2^{\prime}(x) \right ] \Sigma_+,\nonumber \\
p & = & -i \frac{d}{dx}, \ \ W_{1,2}^{\prime}(x) = \frac{d W_{1,2}(x)}{d x},
\eea
\noindent where $W_1, W_2$ are two arbitrary functions. It may be noted that
$Q_1$ is not the adjoint of $Q_2$ in ${\cal{H}}_D$. The Hamiltonian
that may be constructed in terms of a quadratic form of these supercharges
is also non-hermitian in ${\cal{H}}_D$,
\bea
H & = & \{Q_1, Q_2 \}\nonumber \\
& = & \Pi^2 + \left ( W_+^{\prime} \right )^2 + W_+^{\prime \prime}
\sum_{b=1}^3 R^{3b} \sigma^b\nonumber \\
W_{\pm} & = & \frac{1}{2} \left [ W_1(x) \pm  W_2(x) \right ],\ \
W_{\pm}^{\prime} = \frac{d W_{\pm}}{d x}, \ \
\Pi = p + i W_-^{\prime}. 
\label{sp}
\eea
\noindent The generalized momentum operator $\Pi$ includes an imaginary gauge
potential corresponding to the real part of $W_-^{\prime}$. It may be noted
that such imaginary gauge potentials are relevant in the context of
metal-insulator transitions or depinnning of flux lines from extended defects
in type-II superconductors\cite{hn}. The imaginary gauge potential also appears
in the study of unzipping of DNA\cite{somen}. 
Apart from the imaginary gauge potential containing in $W_-^{\prime}$, 
the non-Dirac-hermiticity in $H$ is also introduced through the complex
parameters $R^{3b}$ and complex functions $W_{1,2}$. The appearance of $R^{3b}$
is due to a non-standard non-Dirac-hermitian representation of the Pauli
matrices which may be interpreted as arising due to a gauge-transformation
with complex gauge potentials.

The Hamiltonian is not in the diagonal form due to a non-standard
representation of the Pauli matrices. The eigen-value equation of
the non-Dirac-hermitian $H$ is thus given in terms of a set of two coupled
second-order differential equations. Such coupled differential equations appear
in a variety of physical situations\cite{das}.
The Hamiltonian can be brought to a diagonal form by defining,
\bea
&&\eta_+  :=  \eta_+^b \otimes \eta_+^f, \ \
\eta_+^b := e^{-2 Re(W_-)}\nonumber \\
&&\rho  :=  \rho^b \otimes \rho^f, \ \
\rho^b := e^{- Re(W_-)}, \ \ U:= e^{-i Im(W_-)},
\eea
\noindent and using the similarity transformation,
\bea
h & := & \left ( U \rho \right ) H \left ( U \rho \right )^{-1}\nonumber \\
& = & p^2 + \left ( W_+^{\prime} \right )^2 +
W_+^{\prime \prime} \sigma^3, \ \
\eea
\noindent where $Re(f)/Im(f)$ corresponds to real/imaginary part of $f$.
The complex gauge potential has also been removed by the combined use of
the similarity operator $\rho^b$ and the unitary operator $U$.

In general, the Hamiltonian $h$ is non-Dirac-hermitian due to the
appearance of complex $W_+^{\prime}$ and $W_+^{\prime \prime}$. A number of
supersymmetric systems with complex superpotentials $W_+^{\prime}$ have been
shown to admit entirely real spectra\cite{mostafa1,quesne,quesne3,quesne2,
pani,andrianov,me}.
However, the metric or the inner-product in the Hilbert space is not known
for most of the cases. There may exist similarity transformations which map
$h$ to a Dirac-hermitian Hamiltonian for specific choices of $W_1$ and $W_2$.
However, such an investigation is beyond the scope of this article and
henceforth, $W_+$ is considered to be real so that $h$ is Dirac-hermitian.
It may be noted that real $W_+$ can be obtained even for complex $W_1$ and
$W_2$. In particular, $W_1$ and $W_2$ may be chosen as,
\be
W_1(x) = W(x) + \chi(x) + i \theta_1(x), \ \ W_2(x)= W(x) - \chi(x) +
i \theta_2(x),
\label{deco}
\ee
\noindent where $W(x), \chi(x), \theta_1(x)$ and $\theta_2(x)$ are four
independent real functions of their argument. A real $W_+=W$ is obtained for
$\theta_1=-\theta_2\equiv \theta$. Further, with the choice of $W$ as any
shape-invariant superpotential of Dirac-hermitian quantum system\cite{khare},
the non-Dirac-hermitian Hamiltonian $H$ becomes exactly solvable with entirely
real spectra and unitary time evolution in ${\cal{H}}_{\eta_+}$. It may be
noted that the metric $\eta_+^b=e^{-2 \chi}$ solely depends on the choice of
$\chi$ and does not change for different choices of $W, \theta_1,\theta_2$.
Consequently, appropriate choices of $\chi$ would result in positive-definite
and bounded metric.

The eigenfunctions $\Phi_n$ of $h$ with associated real eigenvalues $E_n$,
satisfying the eigenvalue equation,
\be
h \Phi_n = E_n \Phi_n, \ \
\Phi_n \equiv \pmatrix{ {\Phi_n^+}\cr\\ {\Phi_n^-} },
\ee
\noindent may be used to construct the eigenfunctions of $H$. In particular,
the eigenfunctions $\chi_n$ of $H$ with eigenvalues $E_n$ are,
\bea
\chi_n & = & \left ( U \rho \right )^{-1} \Phi_n\nonumber \\
& = & e^{\chi(x) + i \theta(x)} \pmatrix{ {\Delta^+ \ \Phi_n^+ + 
n_- sinh (\frac{\phi}{2}) \ \Phi_n^-}\cr
{ n_+ sinh (\frac{\phi}{2}) \Phi_n^+ + \Delta^-\Phi_n^-
}},\nonumber \\
\Delta^{\pm} & = & cosh (\frac{\phi}{2}) \pm n^3
sinh(\frac{\phi}{2}), \ \ n_{\pm} = n^1 \pm i n^2.
\eea
\noindent  It may be noted that $\Phi_n$ constitute a complete set of
orthonormal eigenfunctions in ${\cal{H}}_D$, while $\chi_n$ are a complete
set of orthonormal eigenstates in ${\cal{H}}_{\eta_+}$.

\section{Many-particle non-Dirac-hermitian SUSY}

The metric $\eta_+$ is chosen as,
\be
\eta_+^b := e^{-2 \left ( \delta \hat{B} + Re(W_-) \right ) }, \ \
\eta_+:= \eta_+^b \otimes \eta_+^f, \ \ \delta \in R,
\label{metric}
\ee
\noindent where $\eta_+^f$ is given by Eq. (\ref{met}) and $W_{\pm}$ is as
defined in Eq. (\ref{sp}) with the understanding that $W_1, W_2$ are now
functions of the $N$ bosonic co-ordinates. The operator $\hat{B}$
acts on the bosonic co-ordinates only. Specific choice of $\hat{B}$ will be
made in the next section while presenting a few examples. It is assumed that
the bosonic co-ordinates $x_i$ and the momenta $p_i$ are not hermitian in
${\cal{H}}_{\eta_+}$ for the type of operator $\hat{B}$ that will be considered
in this article. A set of hermitian co-ordinates $X_i$ and momenta $P_i$
in ${\cal{H}}_{\eta_+}$ may be introduced as follows,
\be
X_i = \rho^{-1} x_i \rho, \ \ P_i= \rho^{-1} p_i \rho, \ \
\rho:=\sqrt{\eta_+}.
\label{inv}
\ee
\noindent Although the operators $X_i, P_i$ are non-Dirac-hermitian, they
satisfy the basic canonical commutation relations $\left [X_i, P_j \right ]
= \delta_{ij}$. Further, the length in the momentum space as well as
in the co-ordinate space remains invariant under the transformations defined
by the Eq. (\ref{inv}).

The supercharges are introduced as follows,
\bea
&& \tilde{Q}_1 = \sum_{i=1}^N e^{-\gamma_i} \ \psi_i  \left ( P_i +
i W_{1,i} \right )\nonumber \\
&& \tilde{Q}_2 = \sum_{i=1}^N e^{\gamma_i} \ \psi_i^{\dagger} \left ( P_i -
i W_{2,i} \right ),\nonumber \\
&& P_i = - i \frac{\partial}{\partial X_i},
\ \ W_{1,i} = \frac{\partial W_{1}} {\partial X_i}, \ \
\ \ W_{2,i} = \frac{\partial W_{2}} {\partial X_i}, \ \
\gamma_i \in R \ \forall \ i,
\eea
\noindent which are not adjoint of each other in ${\cal{H}}_D$.
The supersymmetric Hamiltonian that may be constructed by using
these supercharges reads,
\bea
\tilde{H} & := & \{Q_1, Q_2\}\nonumber \\
& = & \sum_{i=1}^N \left [ \Pi_i^2 + \left ( W_{+,i} \right )^2 -
W_{+,ii} \right ] + 2 \sum_{i,j=1}^N e^{\gamma_i-\gamma_j} W_{+,ij}
\psi_i^\dagger \psi_j,\nonumber \\
W_{\pm,i} & \equiv & \frac{1}{2} \left ( W_{1,i} \pm W_{2,i} \right ), \ \
\Pi_i := P_i +i W_{-,i}, \ \ 
W_{+,ij} \equiv \frac{\partial^2 W_+}{\partial X_i \partial X_j}. \
\eea
\noindent In general, the Hamiltonian $\tilde{H}$ is non-Dirac-hermitian.
Apart from the complex superpotentials $W_1, W_2$, the non-Dirac-hermitian
interactions are introduced in $\tilde{H}$ through imaginary gauge potentials
containing in the generalized momentum operators $\Pi_i$ and fermion operators
$\Psi_i$. It is worth re-emphasizing that imaginary gauge potentials appear in
the study of a diverse branches of physics including metal-insulator
transitions or depinning of flux lines from extended defects in type-II
superconductors\cite{hn} and unzipping of DNA\cite{somen}. The
non-Dirac-hermitian bosonic potentials may appear in $\tilde{H}$ depending
on specific physical situations and a few such explicit examples will be
discussed in the next section. 

The decomposition of $W_1, W_2$ in Eq. (\ref{deco}) is used in this section
with the understanding that $W, \chi, \theta_1, \theta_2$ are real functions
of the $N$ bosonic co-ordinates. The metric $\eta_+^b$, the operators $\Pi_i$
and the functions $W_{+,i}$ may be re-written as,
\bea
&& \eta_+^b = e^{- 2 (\delta \hat{B} + \chi)}, \ \
\Pi_i = P_i + i \chi_i -\frac{1}{2} \left (\theta_{1,i}-\theta_{2,i}
\right ),\nonumber \\
&& W_{+,i}= W_i + \frac{i}{2} \left (\theta_{1,i} + \theta_{2,i} \right ), \ \
F_i = \frac{\partial F}{\partial X_i}, \ \ F \equiv \{ W, \chi, \theta_1,
\theta_2 \}.
\eea
\noindent The Hamiltonian $\tilde{H}$ becomes hermitian in the Hilbert space
${\cal{H}}_{\eta_+}$ provided the following condition is satisfied,
\be
\theta_1 = - \theta_2 \equiv \theta.
\label{e34}
\ee
\noindent The hermiticity can be checked by re-expressing $\tilde{H}$ as,
\be
\tilde{H} = \sum_{i=1}^N \left [ \Pi_i^2 
+ \left ( W_{i} \right )^2 - W_{ii} \right ] + 
2 \sum_{i,j=1}^N W_{ij} \Psi_i^\dagger \Psi_j.
\ee
\noindent It may be noted that the non-Dirac-hermitian operators $\Pi_i=
p_i + i \chi_i - \theta_{i} $ and non-Dirac-hermitian functions $W_i, W_{ij}$
are hermitian in ${\cal{H}}_{\eta_+}$.

The Hamiltonian $\tilde{H}$ is isospectral with the Dirac-hermitian
Hamiltonian $H$,
\bea
H & := & \left (e^{-i \theta} \rho \right) \tilde{H} \left (e^{-i \theta} \rho
\right)^{-1},\nonumber \\
& = & e^{-i \theta} \left \{ \sum_{i=1}^N \left [ \left ( p_i - \theta_i 
\right )^2 + \left ( \frac{\partial W}{\partial x_i} \right )^2 -
\frac{\partial^2 W}{\partial x_i^2} \right ] + 2 \sum_{i,j=1}^N 
\frac{\partial^2 W}{\partial x_i x_j} \psi_i^\dagger \psi_j \right \}
e^{i \theta},\nonumber \\
& = & \sum_{i=1}^N \left [ p_i^2 + \left ( \frac{\partial W}{\partial x_i}
\right )^2 - \frac{\partial^2 W}{\partial x_i^2} \right ] +
2 \sum_{i,j=1}^N \frac{\partial^2 W}{\partial x_i x_j} \psi_i^\dagger \psi_j.
\eea
\noindent Specific choices of $W$ for which exactly solvable many-particle
supersymmetric quantum systems $H$ are known may be used to construct
iso-spectral non-Dirac-hermitian quantum systems $\tilde{H}$. A set of
orthonormal eigenfunctions $\chi_n$ of $\tilde{H}$ in ${\cal{H}}_{\eta_+}$
may be constructed from the orthonormal eigenfunctions $\Phi_n$ of $H$ in
$H_D$ by using the relation, $\chi_n = (U\rho)^{-1} \Phi_n$.

\section{Examples}

A few specific examples of non-Dirac-hermitian supersymmetric systems with
complex bosonic potentials are considered in this section.
The operator $\hat{B}$ appearing in the metric $\eta_+^b$ in Eq.
(\ref{metric}) is chosen as,
\be
\hat{B} := {\cal{L}}_{12} = x_1 p_2 - x_2 p_1.
\ee
\noindent The presence of ${\cal{L}}_{12}$ in the metric allows
non-Dirac-hermitian bosonic potential in the Hamiltonian\cite{piju1}. In
particular, the co-ordinates $x_1, x_2$ and the momenta $p_1, p_2$ are not
hermitian in ${\cal{H}}_{\eta_+^b}$. A new set of canonical conjugate operators
those are hermitian in the Hilbert space ${\cal{H}_{\eta_+}}$ may be introduced
by using the relation (\ref{inv}) as follows\cite{piju1,ijtp}:
\bea
&& X_1 = x_1 \ cosh \delta + i x_2 \ sinh \delta,\nonumber \\
&& X_2 = - i x_1 \ sinh \delta
+ x_2 \ cosh \delta, \ \ X_i=x_i \ for \ i > 2\nonumber \\
&& P_1 = p_1 \ cosh \delta +  i p_2 \ sinh \delta,\nonumber \\
&& P_2 = - i p_1 \ sinh \delta +
p_2 \ cosh \delta, \ P_i= p_i \ for \ i > 2.
\label{newcor}
\eea
\noindent It may be noted that $L_{12}= X_1 P_2 - X_2 P_1={\cal{L}}_{12}$
is hermitian both in ${\cal{H}}_D$ and ${\cal{H}}_{\eta_+}$. This ensures
that $\eta_+$ defined in Eq. (\ref{metric}) is positive-definite.
Without loss of any generality, $\theta$ is chosen as zero, since
it can always be rotated away by using the unitary operator $U:=e^{-i \theta}$.
The generalized momentum operators $\Pi_i$ now reads,
\be
\Pi_i=P_i + i \chi_i.
\ee
\noindent The imaginary gauge potentials $\chi_i$ can be removed from $\Pi_i$
by using a non-unitary similarity transformation. However, it should be noted
that $P_i$ and $\Pi_i$ act on different Hilbert spaces.

An anti-linear ${\cal{PT}}$ transformation for the bosonic coordinates may
be introduced as follows\cite{ijtp}:
\bea
&& {\cal{P}}: x_1 \leftrightarrow x_2, \ p_1 \leftrightarrow p_2, \
(x_i, p_i) \rightarrow (x_i,p_i) \ \forall \ i > 2;\nonumber \\
&& {\cal{T}}: i \rightarrow - i, \ x_i \rightarrow x_i, \ \
p_i \rightarrow - p_i;\nonumber \\
&& {\cal{PT}}: X_1 \leftrightarrow X_2, \ P_1 \leftrightarrow -P_2, \
(X_i, P_i) \rightarrow (X_i, -P_i) \ \forall i >2;\nonumber \\
&& {\cal{PT}}: \Pi_1 \leftrightarrow - \Pi_2, \ \Pi_i \rightarrow - \Pi_i \
\forall i > 2.
\label{pteqb}
\eea
\noindent The transformations in the last two lines of Eq. (\ref{pteqb})
are derived from the defining relations in the first two lines. Further,
the transformation of $\Pi_i$ under ${\cal{PT}}$, as stated in the last
line of (\ref{pteqb}), is only valid for ${\cal{P}}$-symmetric $\chi$. 
It may be noted that although $P_1^2(\Pi_1^2)$ or $P_2^2(\Pi_2^2)$ are not
${\cal{PT}}$-symmetric individually, the combinations $P_1^2 + P_2^2$
and $\Pi_1^2 + \Pi_2^2$ are always ${\cal{PT}}$-symmetric.

\subsection{Pauli Hamiltonian}

The supercharge $Q$ is introduced as,
\be
Q = \sum_{a=1}^3 \tilde{\Pi}_a \Sigma^a, \ \ \tilde{\Pi}_a = \Pi_a - A_a,
\ee
\noindent where the vector potential $\vec{A}$ with the components
$A_1, A_2, A_3$ are given by,
\bea
A_1 & = & -\frac{B}{2} X_2 =
\frac{B}{2} \left ( i x_1 sinh \delta - x_2 cosh \delta \right ),\nonumber \\
A_2 & = & \frac{B}{2} X_1 =
\frac{B}{2} \left ( x_1 cosh \delta +i x_2 sinh \delta \right ), \ \
A_3 = 0.
\eea
\noindent Although the vector potential $\vec{A}$ is non-Dirac-hermitian and
contains an imaginary part, the magnitude $B$ of the corresponding magnetic
field is a real constant and points in the $z$-direction. The operators
$\tilde{\Pi}_a$, $\Sigma^a$ and the supercharge $Q$ are hermitian in
${\cal{H}}_{\eta_+}$. A non-Dirac-hermitian Hamiltonian is introduced as,
\be
\tilde{H} := Q^2 = \sum_{a=1}^3 \tilde{\Pi}_a^2 - i B \Sigma^3,
\ee
\noindent which is also ${\cal{PT}}$ symmetric with the transformation under
${\cal{P}}$ as given in Eqs. (\ref{pteqs}) and (\ref{pteqb}), while the
action of ${\cal{T}}$ is defined as,
${\cal{T}}: i \rightarrow - i, x_1 \leftrightarrow -x_2,
p_{1,2} \rightarrow - p_{1,2}, \sigma^{1,2,3} \rightarrow - \sigma^{1,2,3}$.
The Hamiltonian $\tilde{H}$ is isospectral with the non-relativistic
Dirac-hermitian Pauli Hamiltonian\cite{vinet}, which can be shown by
using a similarity transformation with $\rho$ as the similarity operator.

\subsection{Super-conformal Quantum System}

The superpotential $W$ to describe a super-conformal quantum system is
chosen as,
\be
W = - \lambda \ ln r, \ \ r= \left ( \sum_{i=1}^N x_i^2 \right )^{\frac{1}{2}}
= \left ( \sum_{i=1}^N X_i^2 \right )^{\frac{1}{2}},
\ee
\noindent which is Dirac-hermitian as well as hermitian in
${\cal{H}}_{\eta_+}$. With this choice of the superpotential,
\be
W_i = - \lambda \frac{X_i}{r^2}, \ \
W_{ij} = \frac{\lambda}{r^4} \left ( 2 X_i X_j - \delta_{ij} r^2 \right ), \ \
\sum_{i=1}^N W_{ii} = - \frac{\lambda (N-2)}{r^2},
\ee
\noindent and the non-Dirac-hermitian supersymmetric Hamiltonian reads,
\be
\tilde{H}=  \sum_{i=1}^N \Pi_i^2 +
\frac{\lambda(\lambda+N-2)}{r^2}
+ \frac{2 \lambda}{r^4} \sum_{i,j=1}^N e^{\gamma_i - \gamma_j}
\left ( 2 X_i X_j - \delta_{ij} r^2 \right ) \psi_i^{\dagger} \psi_j.
\ee
\noindent The purely bosonic Hamiltonian is obtained by projecting $\tilde{H}$
in the fermionic vacuum $|0\rangle_{\eta_+}$. The resulting Hamiltonian with
a further choice of $\chi=0$ is Dirac-hermitian and has been studied\cite{dff}
extensively as a model of conformal quantum system. The Hamiltonian
$\tilde{H}$ reduces to the super-conformal quantum Hamiltonian\cite{fr} in
the limit $\phi=\chi=0, \gamma_i= 0 \ \forall i$. The Hamiltonian $\tilde{H}$
is invariant under a combined ${\cal{PT}}$ operation as defined in
Eqs. (\ref{pteqf}) and (\ref{pteqb}).

The Hamiltonian $\tilde{H}$ along with $D$ and $K$,
\be
D = -\frac{1}{4} \sum_{i=1}^N \left ( X_i \Pi_i + \Pi_i X_i \right ), \ \
K = \frac{1}{4} \sum_{i=1}^N X_i^2,
\ee
\noindent satisfy the $O(2,1)$ algebra which appears as a bosonic sub-algebra
of the $OSp(2|2)$ super-group. The complete algebra of the $OSp(2|2)$ can be
realized by defining,
\be
S_1= \frac{1}{{2}} \sum_{i=1}^N e^{-\gamma_i} \psi_i X_i, \ \
S_2= \frac{1}{{2}} \sum_{i=1}^N e^{\gamma_i} \psi_i^{\dagger} X_i, \ \
Y = \frac{1}{4} \sum_{i=1}^N \left [ \psi_i^{\dagger}, \psi_i \right ].
\ee
\noindent The non-Dirac-hermitian generators $\tilde{H}$, $D$, $S_1$ and $S_2$
of $OSp(2|2)$ are hermitian in ${\cal{H}}_{\eta_+}$. The Dirac-hermitian
generators $K$ and $Y$ are also hermitian in ${\cal{H}}_{\eta_+}$.

The zero energy ground-state wave-function of $\tilde{H}$,
\be
\psi_0 = e^{\chi} r^{\lambda} \ |0\rangle_{{\cal{H}}_{\eta_+}},
\ee
\noindent is not even plane-wave normalizable in ${\cal{H}}_{\eta_+}$
for any choices of $\chi$. Following the prescription\cite{dff,fr}, a compact
operator of the sub-group $O(2,1) \times U(1)$ of $OSp(2|2)$ may be chosen
to study the time-evolution of the system. The relevant non-Dirac hermitian
Hamiltonian,
\be
H_{\pm}^{\prime} = \tilde{H} + K \pm Y.
\ee
\noindent has a complete description including entirely real spectra and
unitary time-evolution in ${\cal{H}}_{\eta_+}$. In fact, the Hamiltonian
$H^{\prime}$ is isospectral with the supersymmetric quantum system
with inverse-square and harmonic potentials\cite{fr}. It is worth mentioning
here that the dynamical supersymmetry of $\tilde{H}$ is $SU(1,1|2)$ for $N=2$
with the following non-Dirac-hermitian realization of the $SU(2)$ generators,
\be
J_1:= e^{-(\gamma_1+\gamma_2)} \psi_1 \psi_2, \\
J_2:=e^{\gamma_1+\gamma_2} \psi_1^{\dagger} \psi_2^{\dagger}, \ \
J_3:= \psi_1^{\dagger} \psi_1 + \psi_2^{\dagger} \psi_2 -1.
\ee
\noindent The relevant discussions for a Dirac-hermitian system\cite{fr,cmm}
may be generalized for the non-Dirac-hermitian Hamiltonian $\tilde{H}$ in a
straightforward way.

A comment is in order before the end of this section.
Normalizable zero energy eigenfunctions of $\tilde{H}$ in ${\cal{H}}_D$
exist for specific choices of $\chi$. For example, with the choice of
$\chi= - \frac{\kappa^2}{2} r^2, \kappa \in R$, the normalizable zero energy
eigenfunctions of $\tilde{H}$ in ${\cal{H}}_D$ are,
\bea
\tilde{\psi}_1 & = & e^{-\frac{\kappa^2}{2} r^2 } r^{\lambda}
\ |0\rangle_{{\cal{H}}_{D}}, \ \ \lambda > 0,\nonumber \\
\tilde{\psi}_2 & = & e^{-\frac{\kappa^2}{2} r^2 } r^{-\lambda}
\ |N\rangle_{{\cal{H}}_{D}}, \ \ \lambda < 0,
\eea
\noindent where $|N\rangle_{{\cal{H}}_D}$ is the conjugate vacuum satisfying
$\psi_i^{\dagger} |N\rangle_{{\cal{H}}_{D}} = 0 \ \forall i$. The supersymmetry
is preserved for the entire range of the parameter $\lambda$. It may be noted
that a scale is introduced in the system for $\kappa \neq 0$.

\subsection{Calogero-type systems}

The superpotential $W$ is chosen as,
\be
W(X_1, X_2, \dots, X_N) = - ln G(X_1, X_2, \dots, X_N) +
\frac{1}{2} \sum_{i=1}^N X_i^2,
\ee
\noindent where $G$ is a homogeneous function of degree
$d$,
\bea
&& \sum_{i=1}^N X_i \frac{\partial G(X_1, \dots, X_N)}{\partial X_i} =
d G(X_1, \dots, X_N),\nonumber \\
&& \sum_{i=1}^N x_i \frac{\partial G(x_1, \dots x_N)}{\partial x_i} =
d G(x_1, \dots, x_N).
\eea
\noindent The homogeneity condition on $G$ is to ensure that the many-body
interaction scales inverse-squarely\cite{pcalo}. Rational Calogero-type models
corresponding to different root systems may be introduced for specific choices
of $G$\cite{pcalo}. The $A_{N+1}$ Calogero-type model is obtained for the
choice,
\be
G = \prod_{i<j=1}^N \left ( X_i - X_j \right )^{\lambda}.
\ee
\noindent The non-Dirac-hermitian Hamiltonian reads,
\bea
\tilde{H} & = & \sum_{i=1}^N \Pi_i^2
+ \lambda (\lambda-1) \sum_{i \neq j=1}^N X_{ij}^{-2} +
\sum_{i=1}^N  x_i^2\nonumber \\
& + & 2 \lambda \sum_{i \neq j=1}^N X_{ij}^{-2} \left ( \psi_i^{\dagger} \psi_i
- e^{\gamma_i-\gamma_j} \psi_i^{\dagger} \psi_j \right )+
2 \sum_{i=1}^N \psi_i^{\dagger} \psi_i
-N - \lambda N (N-1),\nonumber \\
X_{12} & = & \left ( x_1 - x_2 \right ) cosh \delta + i \left ( x_1 +
x_2 \right ) sinh \delta,\nonumber \\
X_{1j} & = &  x_1 cosh \delta + i x_2 sinh \delta - x_j, \ \ j > 2,\nonumber \\
X_{2 j} & = &  -i x_1 sinh \delta + x_2 cosh \delta - x_j \ \ j > 2,\nonumber \\
X_{ij} & = & x_i - x_j, \ \  (i, j) > 2.
\label{calo}
\eea
\noindent Unlike the rational Calogero model\cite{csm}, the many-body
inverse-square interaction term in $\tilde{H}$ is neither invariant
under translation nor singular for $x_{1} = x_i, i > 1$ and  $x_2 = x_i, i > 2$.
Further, the permutation of the bosonic and fermionic coordinates
$x_i \leftrightarrow x_j, \psi_i \leftrightarrow \psi_j, \psi_i^{\dagger}
\leftrightarrow \psi_j^{\dagger}$ does not keep $\tilde{H}$ invariant.
However, the Hamiltonian is invariant under a combined ${\cal{PT}}$ operation
as defined in Eqs. (\ref{pteqf}) and (\ref{pteqb}).

The Hamiltonian $\tilde{H}$ can be mapped to the Dirac-hermitian rational
$A_{N+1}$ Calogero model through a similarity transformation and thus, these
models are isospectral. A word of caution is in order at this point. The
rational $A_{N+1}$ Calogero model has been solved for boundary conditions
by both excluding\cite{csm} and including\cite{sae} the singular points from
the configuration space. The Hamiltonian $\tilde{H}$ has $(N-3)(N-2)$ number
of less singular points compared to the standard Calogero model\cite{csm} due
to the non-singular points at $x_{1} = x_i, i > 1$ and  $x_2 = x_i, i > 2$.
Thus, identical boundary conditions should be used for both the
non-Dirac-hermitian and Dirac-hermitian systems in order to claim that these
systems are isospectral.

\section{Conclusions}

A general prescription to construct non-Dirac-hermitian supersymmetric
quantum system that is isospectral with a Dirac-hermitian Hamiltonian has
bee given. The basic canonical (anti-)commutation relations defining the
supersymmetric system have been realized in terms of non-Dirac-hermitian
operators those are hermitian in a Hilbert space that is endowed with a
pre-determined positive-definite metric. The canonical relations involving
bosonic degrees of freedom have been realized following the method described
in Ref. \cite{piju1}. A pseudo-hermitian realization of the Clifford algebra
has been given which has been used to construct supersymmetric quantum
systems. It has been shown that exactly solvable non-Dirac-hermitian
supersymmetric quantum systems those are isospectral with known exactly
solvable Dirac-hermitian system can always be constructed. Specific
examples of non-Dirac-hermitian nonrelativistic Pauli Hamiltonian,
superconformal quantum system and supersymmetric Calogero-type
models have been presented.

The Pauli matrices appear in diverse branches of physics. The pseudo-hermitian
realization of these matrices may be used to construct non-Dirac-hermitian
quantum systems admitting entirely real spectra. Some of the possibilities
include more general pseudo-hermitian spin-chains, Dicke models, random matrix
models etc. Further, the fermionic operators constructed in this article may
be interpreted as standard fermionic operators with imaginary gauge potentials
and may have applications in condensed matter systems.


\addcontentsline{toc}{section}{References}
\begin{thebibliography}{99}

\bibitem{bend} C. M. Bender, Contemp. Phys. {\bf 46}, 277(2005);
C.M. Bender and S. Boettcher, Phys. Rev. Lett. {\bf 80}, 5243(1998).

\bibitem{ali} A. Mostafazadeh, arXiv:0810.5643; A. Mostafazadeh, J. Math Phys.
{\bf 43}, 205(2002); {\bf 43}, 2814(2002); {\bf 43}, 3944(2002).

\bibitem{quasi} F. G. Scholtz, H. B. Geyer and F. J. W. Hahne, Ann. Phys.
{\bf 213}, 74 (1992).

\bibitem{mostafa1}
A. Mostafazadeh, Nucl. Phys. {\bf B640}, 419(2002);
J. Phys. {\bf A37}, 10193 (2004).


\bibitem{quesne}
C. Quesne, J. Phys. {\bf A41}, 244022(2008).


\bibitem{quesne3}
B. Bagchi, A. Banerjee, E. Caliceti, F. Cannata, H. B. Geyer, C. Quesne and
M. Znojil, Int. J. Mod. Phys. {\bf A20}, 7107 (2005).


\bibitem{quesne2}
B. Bagchi and C. Quesne, J. Phys. {\bf A43}, 305301(2010);
B. Bagchi, S. Mallik and C. Quesne
Mod. Phys. Lett. {\bf A17}, 1651 (2002);
Int. J. Mod. Phys. {\bf A16}, 2859 (2001);
C. Quesne, B. Bagchi, S. Mallik, H. Bila, V. Jakubsky and M. Znojil,
Czech. J. Phys. {\bf 55}, 1161 (2005).

\bibitem{pani}
K. Abhinav and P. K. Panigrahi
Anns. of Phys. {\bf 325}, 1198(2010)
\bibitem{andrianov}
A. A. Andrianov, F. Cannata and A. V. Sokolov, Nucl. Phys. {\bf B773},
107(2007). 

\bibitem{piju1} P. K. Ghosh, J. Phys. A:Math. Theor. {\bf 43},
125203(2010).

\bibitem{ijtp} 
Constructing exactly solvable pseudo-hermitian many-particle quantum systems
by isospectral deformation,
P. K. Ghosh, arXiv: 1012.0907, To appear in Int. J. Theo. Phys.
(DOI: 10.1007/s10773-010-0618-5).


\bibitem{me} P. K. Ghosh, J. Phys. {A38}, 7313 (2005);
T. Deguchi and P. K. Ghosh, Phys. Rev. {\bf E 80}, 021107(2009);
T. Deguchi, P. K. Ghosh and K. Kudo, Phys.  Rev. {\bf E 80}, 026213 (2009).

\bibitem{piju} T. Deguchi and P. K. Ghosh, J. Phys. {\bf A42}, 475208(2009).


\bibitem{ptcsm} B. Basu-Mallick and A. Kundu, Phys. Rev. {\bf B62},
9927(2000); B. Basu-Mallick, T. Bhattacharya and B. P. Mandal, Mod. Phys.
Lett. {\bf A20}, 543(2005); B. Basu-Mallick and B. P. Mandal, Phys. Lett.
{\bf A284}, 231(2001).

\bibitem{pkumar} P. K. Ghosh and K. S. Gupta, Phys. Lett. {\bf A323}, 29(2004).

\bibitem{europe} M. Znojil and M. Tater, J. Phys. {\bf A34}, 1793(2001);
A. fring, Mod. Phys. Lett. {\bf A21}, 691 (2006); A. Fring and M. Znojil,
J. Phys. {\bf A41}, 194010(2008);  P. E. G. Assis and A. Fring,
J. Phys. A: Math. Theor. {\bf 42}, 425206 (2009); A. Fring and M. Smith,
J. Phys. {\bf A43}, 325201(2010).

\bibitem{xxz} F. C. Alcaraz, M. Droz, M. Henkel and V. Rittenberg,
Ann. Phys. (N.Y.) {\bf 230}, 250 (1994);
B. Derrida, M. R. Evans, V. Hakim and V. Pasquier, J. Phys. A: Math. Gen.
{\bf 26}, 1493 (1993).

\bibitem{dicke} E. T. Jaynes and F. W. Cummings, Proc. IEEE {\bf 51}, 89
(1963); M. Tavis and F. W. Cummings, Phys. Rev. {\bf 170}, 379 (1968); 
R. Gilmore and C. M. Bowdon, J. Math. Phys. {\bf 17}, 1617 (1976).

\bibitem{dff} V. de Alfaro, S. Fubini and G. Furlan, Nuovo Cimento {\bf A34},
569(1976).

\bibitem{fr} S. Fubini and E. Rabinovici, Nucl. Phys. {\bf B245}, 17 (1984).

\bibitem{cmm} P. K. Ghosh, J. Phys. {\bf A34}, 5583 (2001).

\bibitem{csm} F. Calogero, J. Math. Phys. {\bf 10}, 2191(1969); J.
Math. Phys. {\bf 10}, 2197(1969);J. Math. Phys. {\bf 12}, 419(1971).

\bibitem{poly} M. A. Olshanetsky and A. M. Perelomov, Phys. Rep. {\bf 71},
314(1981); ibid {\bf 94}, 6 (1983); A. P. Polychronakos, Les Houches 1998
Lectures, hep-th/9902157.

\bibitem{pcalo} P. K. Ghosh, Nucl. Phys. {\bf B595}, 519(2001);
P. K. Ghosh, Nucl. Phys. {\bf B681}, 359(2004).

\bibitem{sae} B. Basu-Mallick, Pijush K. Ghosh and Kumar S. Gupta,
Phys. Lett. {\bf A311}, 87-92 (2003); B. Basu-Mallick, Pijush K. Ghosh
and Kumar S. Gupta, Nucl. Phys. {\bf B659}, 437-457 (2003); L. Feher, I.
Tsutsui and T. Fuolop, Nucl. Phys. {\bf B715}, 713(2005); N. Yonezawa and
I. Tsutsui, J. Math. Phys. {\bf 47}, 012104(2006).

\bibitem{sing} Q.-hai Wang, S.-zhi Chia and J.-hong Zhang, arXiv:1002.267.

\bibitem{jain} Z. Ahmed and S. R. Jain, Phys. Rev. {\bf E67}, 045106(R)
(2003); {\it ibid. } J. Phys. {\bf A36}, 3349 (2009); {\it ibid.}
Mod. Phys. Lett. {\bf A21}, 331 (2006); S. R. Jain and S. C. L.
Srivastava, Phys. Rev. {\bf E78}, 036213 (2008).

\bibitem{lsm} E. Lieb, T. Schultz and D. Mattis, Annals of Physics
{\bf 16}, 407 (1961).

\bibitem{hemen} J. L. van Hemmen, Z. Physik B-Condensed Matter
{\bf 38}, 271 (1980).

\bibitem{clifford} R. Coquereaux, Phys. Lett. {\bf B115} (1982) 389.

\bibitem{ritten} M. De Crombrugghe and V. Rittenberg, Anns. Phys.,
{\bf 151}, 99 (183).

\bibitem{hn} N. Hatano and D. R. Nelson, Phys. Rev. Lett.{\bf 77},
570 (1997); ibid. {\bf B56}, 8651 (1997).

\bibitem{somen} S. M. Bhattacharjee, J.Phys {\bf A33}, L423(2000);
33,9003(E)(2000)

\bibitem{das} T. K. Das and B. Chakrabarty, J. Phys. {\bf A32}, 2387(1999).
\bibitem{khare}
F. Cooper, A. Khare and U. Sukhatme, Phys. Rept. {\bf 251}, 267(1995).
\bibitem{vinet} E. D'Hoker and L. Vinet, Commun. Math. Phys. {\bf 97},
391(1985).
\end{thebibliography}
\end{document}